\documentclass[pra,aps,twocolumn,superscriptaddress,showpacs]{revtex4-1}

\usepackage{graphicx}

\begin{document}

\title{Photoelectron spectra in an autoionization system interacting with a
neighboring atom}

\author{Jan Pe\v{r}ina Jr.}
\affiliation{Institute of Physics of AS CR, Joint Laboratory of
Optics, 17. listopadu 50a, 772 07 Olomouc, Czech Republic}
\author{Anton\'{i}n Luk\v{s}}
\affiliation{Palack\'{y} University, RCPTM, Joint Laboratory of
Optics, 17. listopadu 12, 771 46 Olomouc, Czech Republic}
\author{Wieslaw Leo\'nski}
\affiliation{Quantum Optics and Engineering Division, Institute of
Physics, University of Zielona G\'ora, Prof.~Z.~Szafrana 4a,
65-516 Zielona G\'ora, Poland}
\author{Vlasta Pe\v{r}inov\'{a}}
\affiliation{Palack\'{y} University, RCPTM, Joint Laboratory of
Optics, 17. listopadu 12, 771 46 Olomouc, Czech Republic}
\email{perinaj@prfnw.upol.cz}

\begin{abstract}
Photoelectron ionization spectra of an autoionization system with
one discrete level interacting with a neighbor two-level atom are
discussed. The formula for long-time ionization spectra is
derived. According to this formula, the spectra can be composed of
up to eight peaks. Moreover, the Fano-like zeros for weak optical
pumping have been identified in these spectra. The conditional
ionization spectra depending on the state of the neighbor atom
exhibit oscillations at the Rabi frequency. Dynamical spectral
zeros occurring once per the Rabi period have been revealed in
these spectra.
\end{abstract}

\pacs{32.80.-t,33.80.Eh,34.20.-b}


\keywords{laser-induced ionization, Fano zeros, quantum
interference resonances, atom-atom interaction}

\maketitle

\section{Introduction}

The problem of ionization of an atom with discrete autoionizing
levels has been addressed many times under various conditions
since the pioneering work by Fano \cite{Fano1961} appeared. In his
contribution, Fano has explained the existence of unpopulated
frequencies in the continuum of ionized states by diagonalizing
the appropriate Hamiltonian. This effect occurs as a consequence
of destructive interference of two quantum ionization paths. This
phenomenon occurring at specific frequencies (or for given
ionization states) is spoken of as the presence of 'Fano zeros'.
Moreover, there might occur a strong narrowing of spectral peaks
in the vicinity of such frequencies by virtue of strong
interference. This effect is sometimes referred to as the
'confluence of bound-free coherences' \cite{Rzazewski1981}. These
effects can be degraded to certain extent by, e.g., spontaneous
emission of radiation \cite{Lewenstein1983,Haus1983,Agarwal1984},
finite pump laser bandwidth \cite{Rzazewski1983} or collisions
\cite{Agarwal1984}. In general, there exist $ n $ Fano zeros in an
autoionization system with $ n $ discrete autoionizing levels
\cite{Fano1961,Leonski1987}. Generalizations including several
mutually non-interacting continua has also been given
\cite{Fano1961,Leonski1988}. The presence of autoionizing levels
also influences ionization dynamics under strong laser pumping
\cite{Lambropoulos1981}. Experimental observation of Fano spectral
zeros has been reported, e.g., in \cite{Journel1993}. The
existence of discrete levels in autoionization systems can lead to
transparency for ultra-short pulses \cite{Paspalakis1999} or
slowing-down of propagating light \cite{Raczynski2006}. The
dynamics of ionization can be even influenced by the Zeno or
anti-Zeno effects \cite{Lewenstein2000}. A generalization to
low-light quantum optical fields including the Fock coherent or
squeezed states has also been given \cite{Leonski1990,Leonski1993}
and it has revealed additional interferences in photoelectron
ionization spectra stemming from the discrete energies of
quantized optical fields. The studied ionization quantum-path
interference effects play an important role in spectroscopy in
explaining asymmetric spectral profiles \cite{Durand2001}. Similar
quantum interference effects can be found in many other fields of
physics, both using mass particles and photons. Among others,
semiconductor hetero-structures or photonic waveguides can be
mentioned \cite{Miroshnichenko2010}. An extended list of
references dealing with autoionization can be found in
\cite{Agarwal1984,Leonski1987,Leonski1988a}.

In this work we continue the investigation of the influence of a
neighbor atom to an ionization system and its long-time
photoelectron ionization spectra. The influence is assumed to have
the form of energy transfer cased, e.g., by the dipole-dipole
interaction. Compared to \cite{PerinaJr2011}, we additionally
assume one discrete bound (autoionizing) state present in the
ionization system. Similarly as in \cite{PerinaJr2011}, the
neighbor atom is modelled as a two-level system that undergoes
Rabi oscillations in a stationary optical field. These
oscillations significantly influence conditional photoelectron
spectra of the autoionization system. In these spectra, the
so-called dynamical zeros occurring once per the Rabi period
\cite{PerinaJr2011} have been found. Frequencies in photoelectron
ionization spectra corresponding to both the Fano-like zeros (for
weak optical pumping) and the dynamical zeros are studied using
both analytical and numerical approaches. Molecular condensates
\cite{Silinsh1994} as well as systems of quantum dots or other
semiconductor hetero-structures \cite{Miroshnichenko2010} are
suitable candidates for the verification of the obtained results.

The paper is organized as follows. Sec.~II brings the model
Hamiltonian, solution of the corresponding Schr\"{o}dinger
equation and formulas for photoelectron ionization spectra. These
spectra in their long-time limit are discussed in Sec.~III. The
frequencies of Fano-like zeros in the spectra are analyzed in
Sec.~IV using the method of canonical transformation. Sec.~V is
devoted to dynamical zeros. Conclusions are drawn in Sec.~VI.
Formulas giving the frequencies of poles of the Lorentzian curves
constituting the photoelectron ionization spectra can be found in
Appendix~A. A method for the determination of frequencies of the
Fano as well as dynamical zeros is developed in Appendix~B.

\section{Quantum model and its photoelectron ionization spectra}

\begin{figure}  
 \includegraphics[scale=0.7]{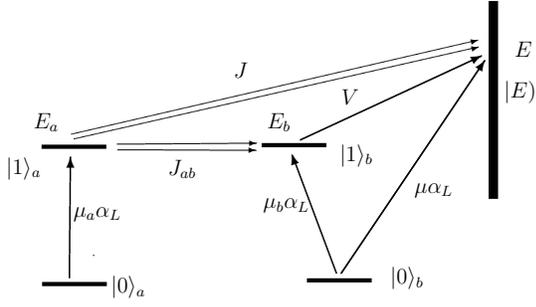}
 \caption{Scheme of the autoionization system $ b $ interacting with a
  two-level atom $ a $. State $ |1\rangle_a $ means an excited state
  of atom $ a $ with the energy $ E_a $, $ |1\rangle_b $ stands for an excited
  bound state of autoionization system $ b $ with the energy $ E_b $, and $ |E) $
  is a free state inside the continuum at the atom $ b $ with the energy $ E $.
  Symbols $ \mu_a $, $ \mu_b $, and $
  \mu $ denote the dipole moments between the ground states $ |0\rangle_a $ and
  $ |0\rangle_b $ and the corresponding excited states, $ \alpha_L $ stands for
  the pumping amplitude,
  $ V $ describes the Coulomb configurational coupling
  between the states $ |1\rangle_b $ and $ |E) $, and $ J $
  [$ J_{ab} $] refers to the dipole-dipole interaction
  between the states $ |1\rangle_a $
  and $ |E) $ [$ |1\rangle_b $]. Double arrows indicate that two electrons at the atoms
  $ a $ and $ b $ participate in the interaction (energy transfer).}
\label{fig1}
\end{figure}

The considered ionization system (atom, molecule) with one
autoionizing level is assumed to interact with a neighbor
two-level atom (molecule) by the dipole-dipole interaction (for
the scheme, see Fig.~\ref{fig1}).
Both the ionization system and the neighbor atom are under the
influence of a stationary optical field. The Hamiltonian $
\hat{H}_{\rm a-i} $ of the ionization system $ b $ with one
autoionizing level can be written in the form ($ \hbar = 1 $ is
assumed, \cite{Meystre2007}):
\begin{eqnarray}   
 \hat{H}_{\rm a-i} &=& E_b|1\rangle_b{}_b\langle1| +
  \int dE  E|E)(E| \nonumber \\
 & & \hspace{-4mm} \mbox{} +
  \int dE \left( V |E)\,{}_b\langle 1|+\mbox{H.c.}
  \right) \nonumber \\
 & & \hspace{-4mm} \mbox{} +
  \left[ \mu_b \alpha_L\exp(-iE_Lt) |1\rangle_b{}_b\langle 0|+\mbox{H.c.}
  \right] \nonumber \\
 & & \hspace{-4mm} \mbox{} +
  \int dE \left[ \mu \alpha_L\exp(-iE_Lt) |E)\,{}_b\langle 0|+\mbox{H.c.}
  \right] .
\label{1}
\end{eqnarray}
In Eq.~(\ref{1}), the symbol $ E_b $ means the excitation energy
from the ground state $ |0\rangle_b $ into the excited bound state
$ |1\rangle_b $ of atom $ b $. The continuum of the autoionization
system $ b $ is formed by the states $ |E) $ with their energies $
E $. The coupling constant $ V $ describes the Coulomb
configuration interaction between the states $ |1\rangle_b $ and $
|E) $ inside the continuum. The dipole moments $ \mu_b $ and $ \mu
$ characterize the optical excitation of the corresponding states;
$ \alpha_L $ means an optical-field amplitude oscillating at the
frequency $ E_L $. The symbol $ \mbox{H.c.} $ stands for the
Hermitian conjugated term.

A neighbor two-level atom $ a $ interacting with the optical field
through the dipole moment $ \mu_a $ is described by the
Jaynes-Cummings Hamiltonian $ \hat{H}_{\rm t-a} $:
\begin{equation}    
 \hat{H}_{\rm t-a} = E_a|1\rangle_a{}_a\langle1|
  + \left[ \mu_a \alpha_L \exp(-iE_Lt) |1\rangle_a{}_a\langle0|+\mbox{H.c.}
  \right] .
\label{2}
\end{equation}
The energy $ E_a $ of the excited bound state $ |1\rangle_a $ of
the atom $ a $ is measured relative to the energy of the ground
state $ |0\rangle_a $. The ground states of atoms $ a $ and $ b $
are assumed to have the same energy that is chosen to be zero.

The energy transfer \cite{Silinsh1994} caused by the dipole-dipole
interaction between the two-level atom $ a $ and the autoionizing
system $ b $ is characterized by the Hamiltonian $ \hat{H}_{\rm
trans}^e $:
\begin{eqnarray}   
 \hat{H}^e_{\rm trans} &=& \left[ J_{ab} |1\rangle_b{}_b\langle0||0\rangle_a{}_a\langle1|
  + \mbox{H.c.} \right] \nonumber \\
 & & \hspace{-1cm} \mbox{} +  \int dE \left[
  J |E)\,{}_b\langle0||0\rangle_a{}_a\langle1|+
  \mbox{H.c.} \right] .
\label{3}
\end{eqnarray}
In this interaction, one electron looses its energy when returning
from the excited state into the ground state, whereas the other
one absorbs this energy and moves from the ground state into its
own excited state. The constants $ J_{ab} $ and $ J $ quantify the
strength of this interaction.

A general quantum state of two electrons at the atoms $ a $ and $
b $ can be written in the following form appropriate to the
rotating frame:
\begin{eqnarray}   
 |\psi\rangle(t) &=& c_{00}(t) |0\rangle_{a} |0\rangle_{b}
   + c_{10}(t) |1\rangle_{a} |0\rangle_{b} \nonumber \\
 & & \hspace{0mm} \mbox{} + c_{01}(t) |0\rangle_{a} |1\rangle_{b}
   + c_{11}(t) |1\rangle_{a} |1\rangle_{b} \nonumber \\
 & & \hspace{0mm} \mbox{}  + \int dE d_{0}(E,t)
  |0\rangle_{a} |E) \nonumber \\
 & & \hspace{0mm} \mbox{} + \int dE
  d_{1}(E,t) |1\rangle_{a} |E).
\label{4}
\end{eqnarray}
The time-dependent coefficients $ c_{00} $, $ c_{10} $, $ c_{01}
$, $ c_{11} $, $ d_0(E) $, and $ d_1(E) $ characterize the state $
|\psi\rangle $ at an arbitrary time.

The Schr\"{o}dinger equation with the Hamiltonian $ \hat{H}_{\rm
a-i} + \hat{H}_{\rm t-a} + \hat{H}^e_{\rm trans} $ can be written
as a system of differential equations for the coefficients of
decomposition written in Eq.~(\ref{4}):
\begin{eqnarray}   
 i \frac{d}{dt} \left[\begin{array}{c}
  {\bf c^e}(t) \\ {\bf d}(E,t) \end{array}\right] =
  \left[\begin{array}{cc}
  {\bf A^e} & {\bf B^e}\int dE \\ {\bf B^{\bf e\dagger}} & {\bf K}(E)
   \end{array} \right]
  \left[\begin{array}{c} {\bf c^e}(t) \\ {\bf d}(E,t)
   \end{array}\right] . \nonumber \\
  & &
\label{5}
\end{eqnarray}
The symbol $ \dagger $ denotes the Hermitian conjugation. The
vectors $ {\bf c^e} $ and $ {\bf d} $ and the matrices $ {\bf A^e}
$, $ {\bf B^e} $, and $ {\bf K} $ introduced in Eq.~(\ref{5}) can
be derived in the form:
\begin{eqnarray}  
 & & {\bf c^e}(t) = \left[\begin{array}{c}
  c_{00}(t) \\ c_{10}(t) \\ c_{01}(t) \\ c_{11}(t)
  \end{array}\right] , \hspace{5mm}
 {\bf d}(E,t) = \left[\begin{array}{c}
  d_{0}(E,t) \\ d_{1}(E,t) \end{array} \right] ,
\label{6}
  \\
 & & {\bf A^e} = \left[\begin{array}{cccc}
  0 & \mu_a^*\alpha_L^* & \mu_b^*\alpha_L^* & 0 \\
  \mu_a\alpha_L & \Delta E_a & J_{ab}^* & \mu_b^*\alpha_L^* \\
  \mu_b\alpha_L & J_{ab} & \Delta E_b & \mu_a^*\alpha_L^* \\
  0 & \mu_b\alpha_L & \mu_a\alpha_L & \Delta E_a + \Delta E_b
  \end{array}\right] , \nonumber \\
 & &
\label{7}
  \\
 & & {\bf B^e} = \left[\begin{array}{cc}
   \mu^*\alpha_L^* & 0 \\ J^* & \mu^*\alpha_L^* \\
   V^* & 0 \\ 0 & V^* \end{array}\right],
\label{8}
   \\
 & & {\bf K}(E) = \left[\begin{array}{cc}
  E-E_L & \mu_a^*\alpha_L^* \\
  \mu_a\alpha_L & E- E_L + \Delta E_a \end{array}\right] ;
\label{9}
\end{eqnarray}
$ \Delta E_a = E_a - E_L $ and $ \Delta E_b = E_b - E_L $. As the
electrons remain inside the system during the evolution, the norm
of state $ |\psi\rangle $ is preserved:
\begin{equation} 
 \sum_{j,k=0}^{1} |c_{jk}(t)|^2 + \sum_{j=0}^{1} \int dE
  |d_j(E,t)|^2 = 1 .
\end{equation}

The system of differential equations (\ref{5}) can be solved using
the Laplace-transform method (for details, see
\cite{PerinaJr2011}). The coefficients $ d_0(E,t) $ and $ d_1(E,t)
$ of the solution then give the amplitude photoelectron ionization
spectra of atom $ b $ conditioned by the presence of atom $ a $ in
the ground and excited states, respectively. It can be shown that,
in the solution, there exist two prominent frequencies $ \xi_1 $
and $ \xi_2 $ characterizing oscillations of the neighbor atom $ a
$;
\begin{eqnarray}   
 \xi_{1,2} &=& E_L -\frac{\Delta E_a \pm \delta\xi }{2} , \nonumber \\
 \delta \xi &=& \sqrt{ (\Delta E_a)^2 + 4|\mu_a\alpha_L|^2 } .
\label{11}
\end{eqnarray}
The solution for the coefficients $ {\bf d}(E,t) $ can be derived
as follows \cite{PerinaJr2011}:
\begin{eqnarray}   
 {\bf d}(E,t) &=& {\bf d}^{\bf \xi_1}(E,t) +
  {\bf d}^{\bf \xi_2}(E,t) ,
\label{12}  \\
 {\bf d}^{\bf \xi_j}(E,t) &=& i {\bf K_{j}} {\bf B^{\bf e\dagger}} {\bf P^e}
  {\bf U_{k}^e}(E,t) {\bf P}^{\bf e-1} {\bf c^e}(0) , \nonumber \\
  & & \mbox{} \hspace{3mm} j=1,2.
\label{13}
\end{eqnarray}
The elements of the diagonal evolution matrices $ {\bf U_{k}^e} $,
$ k=1,2 $, in Eq.~(\ref{13}) are given as:
\begin{eqnarray}   
 \left[ {\bf U_k^e} \right]_{jl}(E,t) &=& \frac{i\delta_{jl}}{E-\Lambda_{M^e,j}-\xi_k}
  \left[ \exp[i(\xi_k-E)t]  \right. \nonumber \\
 & & \hspace{1cm} \left. \mbox{} - \exp(-i\Lambda_{M^e,j}t) \right];
\label{14}
\end{eqnarray}
$ \delta_{jk} $ being the Kronecker symbol. The symbols $
\Lambda_{M^e,j} $ denote eigenvalues of the evolution matrix $
{\bf M^e} $, $ {\bf M^e} = {\bf A^e} - i\pi {\bf B^e} {\bf B^{\bf
e\dagger}} $. The eigenvectors of matrix $ {\bf M^e} $ then form
the columns of matrix $ {\bf P^e} $ introduced in Eq.~(\ref{13}).
The matrices $ {\bf K_k} $ occurring in Eq.~(\ref{13}) take the
form:
\begin{equation}   
 {\bf K_k} = \frac{(-1)^k}{\delta \xi} \left[ \begin{array}{cc}
  E_a+\xi_k & -\mu_a^*\alpha_L^*  \\ -\mu_a\alpha_L &
  E_L+\xi_k \end{array} \right] .
\label{15}
\end{equation}
The vector $ {\bf c^e}(0) $ in Eq.~(\ref{13}) gives the initial
conditions. We assume here that electrons at both the two-level
atom $ a $ and the autoionizing system $ b $ are initially in
their ground states, i.e. $ {\bf c^e}(0) = (1,0,0,0) $.

In deriving the long-time photoelectron ionization spectra, the
long-time form of the evolution matrices $ {\bf U_{k}^e} $ defined
in Eq.~(\ref{14}) is needed:
\begin{equation}   
 \left[ {\bf U_{k}^{\bf e,lt}}\right]_{jl}(E,t) = \frac{i\delta_{jl}
   \exp[i(\xi_k-E)t]}{E-\Lambda_{M^e,j}-\xi_k} .
\label{16}
\end{equation}
We note that the formula in (\ref{16}) is valid for the times $ t
$ obeying $ t \gg 1/|{\rm Im}\{\Lambda_{M^e,j}\}| $ for $
j=1,\ldots,4 $; the symbol $ {\rm Im} $ means the imaginary part.
The long-time form of evolution matrices $ {\bf U_{1}^{e,lt}} $
and $ {\bf U_{2}^{e,lt}} $ shows that oscillations at the Rabi
frequency $ \delta \xi = \xi_1 - \xi_2 $ occur in the intensity
photoelectron ionization spectra $ I_j^{\rm lt} $,  $ I_j^{\rm
lt}(E) \equiv |d_j^{\rm lt}|^2(E) $. A detailed analysis has shown
\cite{PerinaJr2011} that the long-time intensity spectra $
I_0^{\rm lt} $ and $ I_1^{\rm lt} $ can be expressed in a specific
form:
\begin{eqnarray}   
 I^{\rm lt}_0(E,t) &=& I^{\rm st}_0(E) + I^{\rm osc}(E)
  \cos[\delta\xi t+ \varphi(E)] , \nonumber \\
 I^{\rm lt}_1(E,t) &=& I^{\rm st}_1(E) - I^{\rm osc}(E)
  \cos[\delta\xi t+ \varphi(E)] . \nonumber \\
 & &
\label{17}
\end{eqnarray}
The intensities $ I^{\rm st}_0 $ and $ I^{\rm st}_1 $ denote the
steady-state parts of the corresponding spectra, whereas the
intensity $ I^{\rm osc} $ describes the magnitude of harmonic
oscillations between the spectra $ I^{\rm lt}_0 $ and $ I^{\rm
lt}_1 $. The symbol $ \varphi $ stands for a spectrally dependent
phase. These temporal oscillations at the Rabi frequency in the
conditional long-time photoelectron ionization spectra can be
observed using the time-resolved spectroscopy of photo-ionized
electrons \cite{Meystre2007}. If the temporal resolution is not
sufficient, only the steady-state parts $ I^{\rm st}_0(E) $ and $
I^{\rm st}_1(E) $ are experimentally available. We note that the
Rabi oscillations can alternatively be observed in the long-time
behavior of the two-level atom $ a $ provided that a suitable
basis in the continuum of ionized states of atom $ b $ is chosen
and a conditional measurement into its basis functions is
considered.

The form of long-time spectra $ I^{\rm lt}_0 $ and $ I^{\rm lt}_1
$ as written in Eq.~(\ref{17}) guarantees that the overall
long-time photoelectron ionization spectrum $ I^{\rm lt}(E) =
I_0^{\rm lt}(E,t) + I_1^{\rm lt}(E,t) $ is time independent;
\begin{equation}   
 I^{\rm lt}(E) = {I}^{\rm st}_0(E) + {I}^{\rm st}_1(E).
\label{18}
\end{equation}

In the long-time photoelectron ionization spectra, there may occur
frequencies that cannot be populated. If a given frequency $ E_F $
cannot be excited for arbitrary times, we have a Fano zero obeying
the following equation:
\begin{equation}   
 I^{\rm lt}(E_F) = 0 .
\label{19}
\end{equation}
According to Eq.~(\ref{18}) a Fano zero at the frequency $ E_F $
is present only if $ I^{\rm st}_0(E_F) = 0 $ and $ I^{\rm
st}_1(E_F) = 0 $.

It may also happen that the long-time spectral components $ I^{\rm
st}_0 $, $ I^{\rm st}_1 $, and $ I^{\rm osc} $ fulfil one or both
of the following equations for specific frequencies $ E_D $:
\begin{equation}   
 I^{\rm st}_j(E_D) = I^{\rm osc}(E_D), \hspace{5mm} j=0,1.
\label{20}
\end{equation}
This means that the long-time photoelectron ionization spectrum $
I_0^{\rm lt}(E,t_D) $ [or $ I_1^{\rm lt}(E,t_D) $] reaches zero at
suitable time instants $ t_D $. Such a frequency $ E_D $
corresponds to a dynamical zero that periodically occurs with the
Rabi period $ 2\pi/\delta\xi $ \cite{PerinaJr2011}. We note that
dynamical zeros occur, because of the interaction of the
autoionization system $ b $ with the two-level atom $ a $. The
frequencies of dynamical zeros depend, in general, on the state of
the two-level atom $ a $. However, if the two-level atom $ a $ is
resonantly pumped, these frequencies in the long-time
photoelectron spectra $ I_0^{\rm lt} $ and $ I_1^{\rm lt} $
coincide. We note that a Fano zero also obeys the conditions in
Eq.~(\ref{20}) defining a dynamical zero.

\section{Photoelectron ionization spectra}

The long-time photoelectron ionization spectra of the Fano model
as well as the ionization system interacting with a neighbor atom
studied in \cite{PerinaJr2011} are useful in the analysis of the
spectra belonging to the autoionization system interacting with a
neighbor. That is why, we refer to them in the discussion below.

The general form of amplitude photoelectron ionization spectra $
{\bf d}(E,t) $ of the interacting autoionization system is
composed of eight Lorentzian curves as the formulas in
Eqs.~(\ref{12}), (\ref{13}), and (\ref{16}) show. These curves are
located at different frequencies $ E_r $ in the complex plane $ E
$. The complex frequencies $ E_r $ differ in magnitudes of their
complex parts and so they lead to peaks of different widths on the
real axis $ E $. In more detail, there exist two groups of four of
the frequencies $ E_r $ (see Appendix~A). The frequencies $ E_r $
from the second group are just those of the first group shifted by
the Rabi frequency $ \delta\xi $. However, as numerical results
have revealed, only four frequencies $ E_r $ (two pairs) are
important for the determination of shapes of the photoelectron
ionization spectra $ I^{\rm lt} $ in two regimes discussed here
and assuming resonant pumping of atom $ a $.

In order to demonstrate the main features found in the
photoelectron ionization spectra, we first consider the comparable
'ionization' interactions at atoms $ a $ and $ b $ ($ V \approx J
$, $ J_{ab} $). The regime with $ V \gg J $, $ J_{ab} $
appropriate for molecular condensates is analyzed subsequently.

\begin{figure}  
 (a) \includegraphics[scale=0.4]{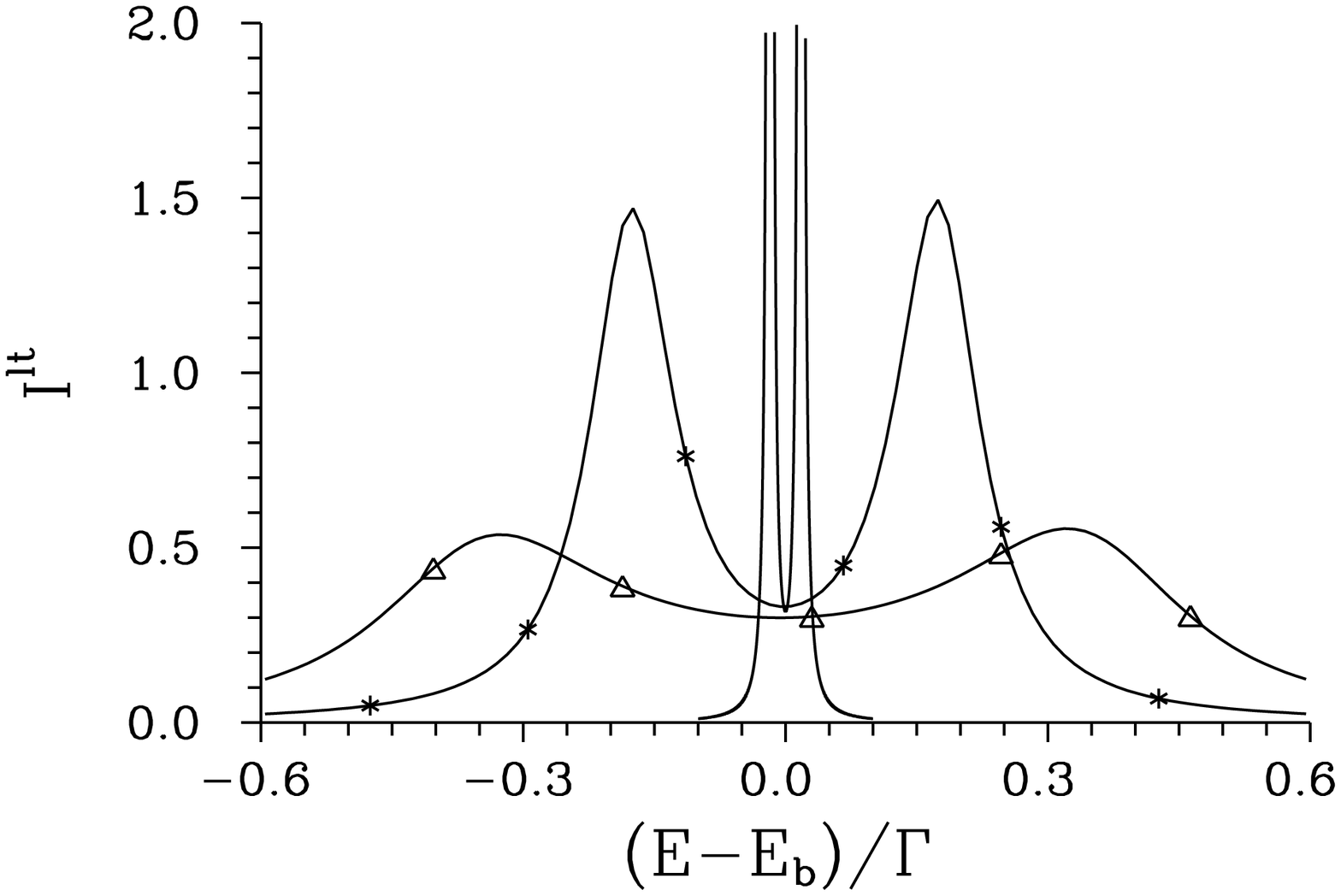}

 \vspace{7mm}
 (b) \includegraphics[scale=0.4]{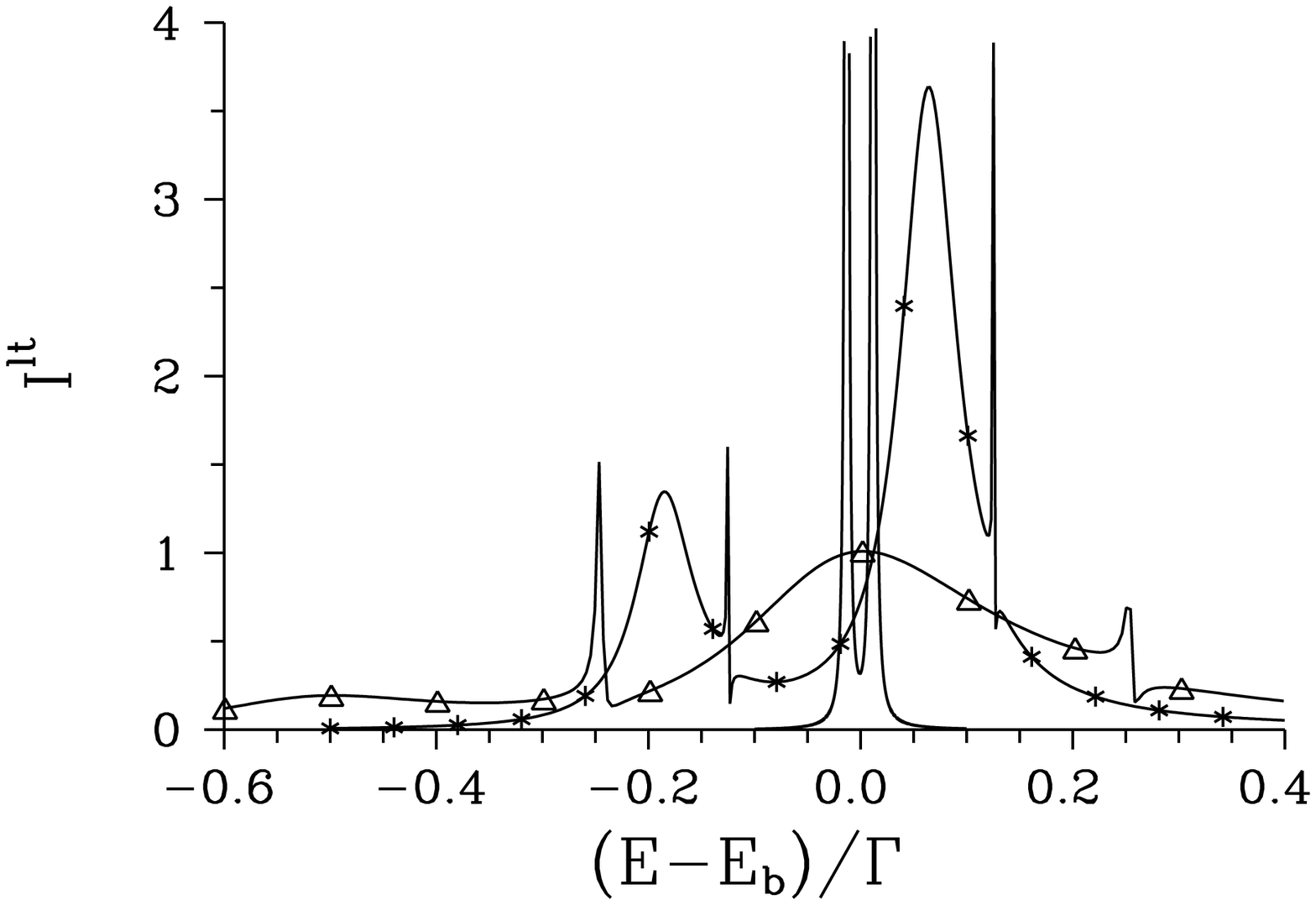}

 \caption{Long-time photoelectron ionization spectra $ I^{\rm lt} $ for
  (a) $ q_a = q_b = 100 $ and (b) $ q_a = q_b = 1 $ for different
  values of pumping parameter $ \Omega $: $ \Omega = 0.1
  $ (solid curve), $ \Omega = 1 $ (solid curve with $ \ast $),
  and $ \Omega = 2 $ (solid curve with $ \triangle $);
  $ \gamma_a = \gamma_b =1 $, $ E_a=E_b=E_L=1 $, $ J_{ab} = 0 $;
  $ q_a = \mu_a/(\pi \mu J^*) $, $ \gamma_a = \pi |J|^2 $,
  $ q_b = \mu_b/(\pi \mu V^*) $, $ \gamma_b = \pi |V|^2 $, $ \Omega =
  \sqrt{4\pi\Gamma} (Q+i)\mu\alpha_L $, $ \Gamma = \gamma_a +
  \gamma_b $, $ Q = (\gamma_a q_a + \gamma_b q_b)/\Gamma $.
  Spectra are normalized such that $ \int dE I^{\rm lt}(E)
  = 1 $.}
\label{fig2}
\end{figure}

\begin{figure}  
 (a) \includegraphics[scale=0.4]{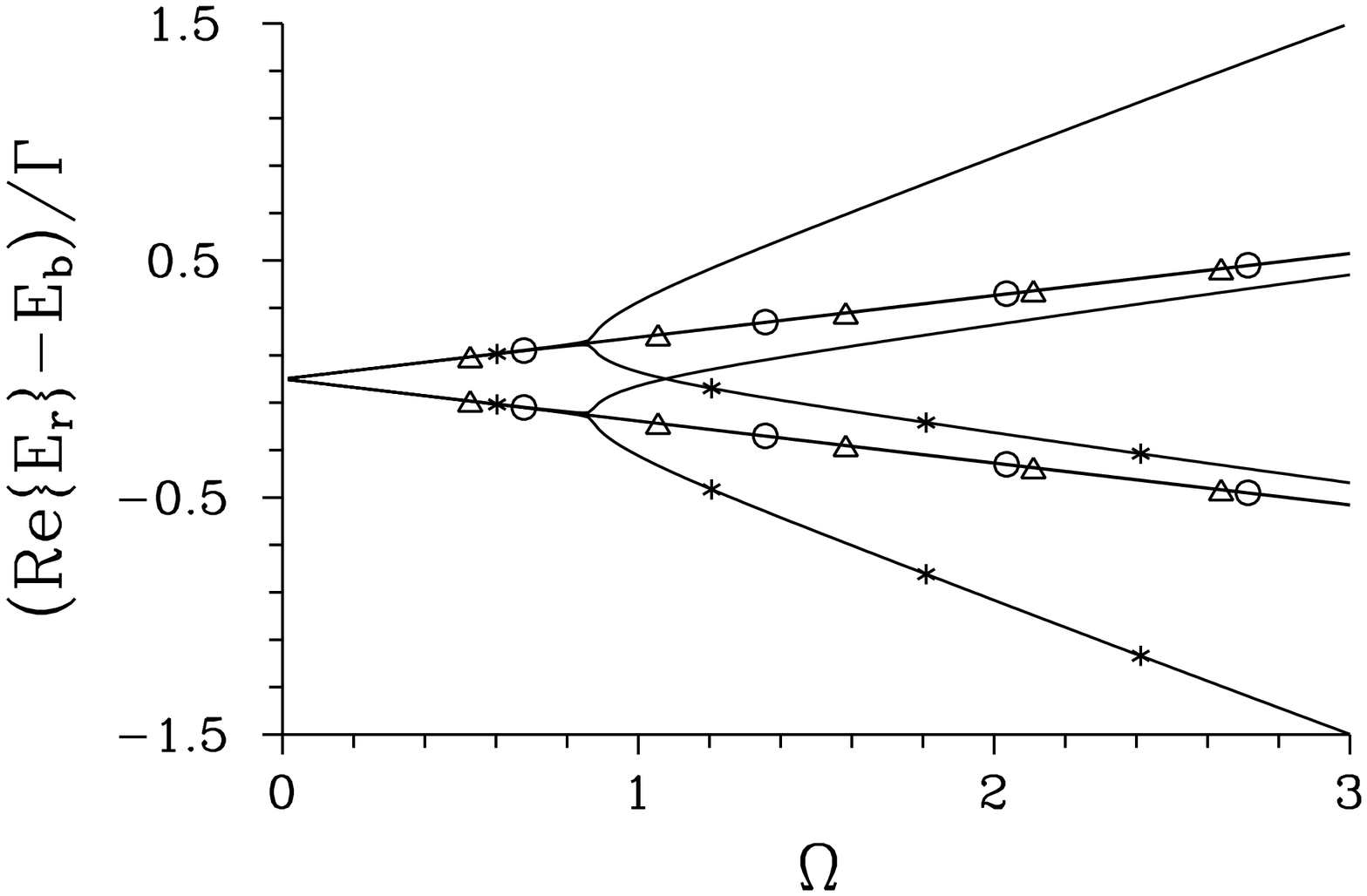}

 \vspace{7mm}
 (b) \includegraphics[scale=0.4]{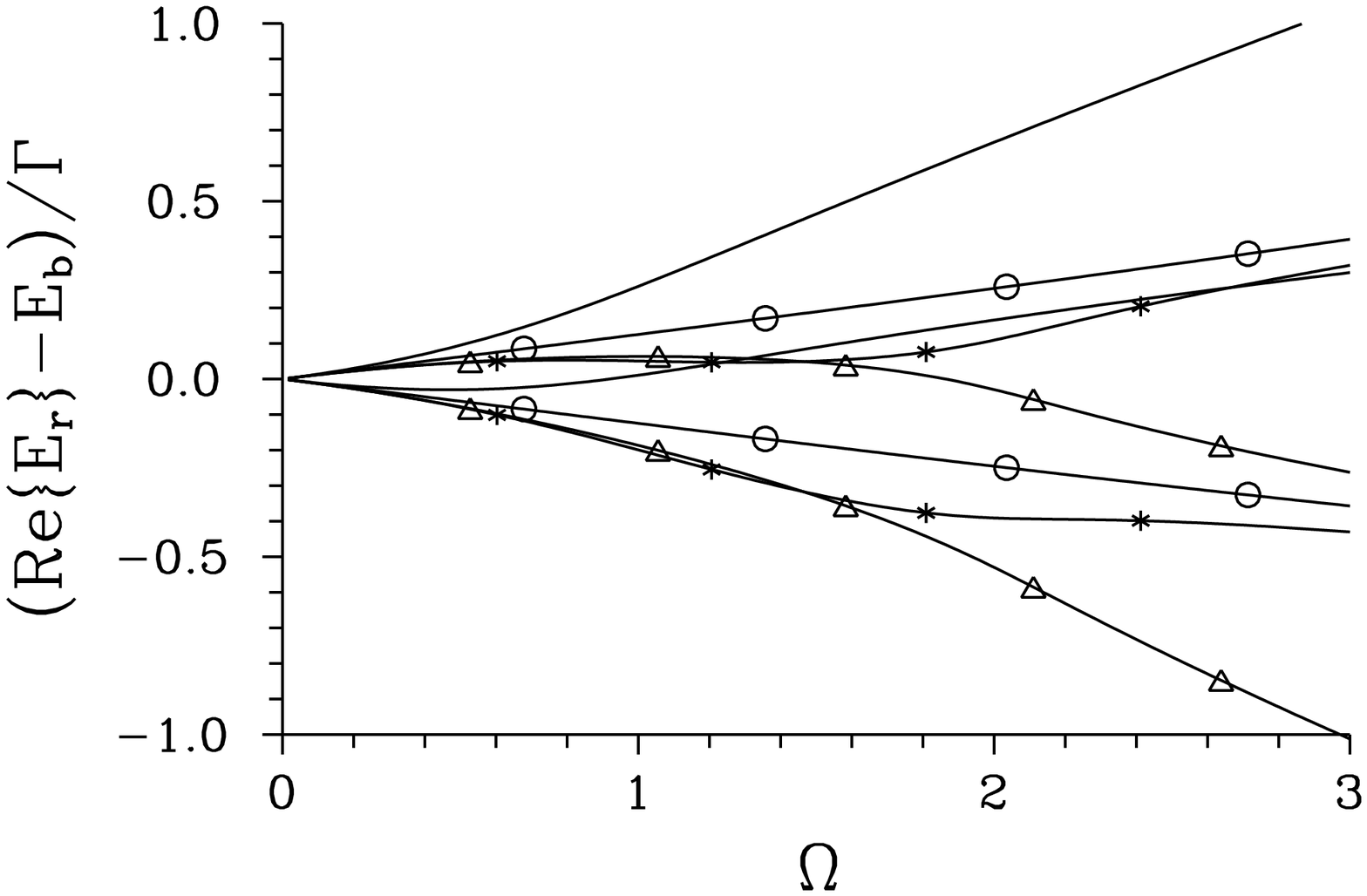}

 \caption{Real parts of eight complex frequencies $ E_r $, $ E_r =
  \Lambda_{M^e,j} + \xi_{k} $ for $ j=1,\ldots, 4 $ and $ k=1,2 $,
  indicating the positions of poles
  in spectra $ I^{\rm lt} $ for
  (a) $ q_a = q_b = 100 $ and (b) $ q_a = q_b = 1 $ as they depend
  on pumping parameter $ \Omega $. Roots are divided into four
  pairs (plotted with different styles of curves) with equal imaginary
  parts and mutual frequency difference $ \delta\xi $.
  Two pairs (solid curves with $ \triangle $ and $ \circ $)
  have small imaginary parts and are thus visible in the spectra  $ I^{\rm lt} $
  shown in Fig.~\ref{fig2}. The values of other parameters are given in
  caption to Fig.~\ref{fig2}.}
\label{fig3}
\end{figure}

Analyzing the long-time photoelectron ionization spectra, we first
consider a weak direct ionization ($ q_a, q_b \gg 1 $, for the
definition of parameters, see the caption to Fig.~\ref{fig2}). On
assuming equally strong indirect ionization paths through the
states $ |1\rangle_a $ and $ |1\rangle_b $ ($ q_a=q_b $), typical
symmetric two-peak photoelectron ionization spectra are observed
[see Fig.~\ref{fig2}(a)]. We note that one should keep in mind
that the ionization process through the state $|1\rangle_a $ is
not typical, since energy transfer without real transfer of
electrons occurs. The greater the pumping parameter $ \Omega $ is,
the larger is the distance between two peaks that form an
Autler-Townes doublet already discussed in the literature about
autoionization processes (see, e.g., \cite{Leonski1988} and
references therein). These conclusions can be drawn from the
positions of eight frequencies $ E_r $ in the complex plane $ E $
discussed above. Their real parts are plotted in
Fig.~\ref{fig3}(a) as a function of the pumping parameter $ \Omega
$. They create pairs with equal imaginary parts and mutual
frequency difference equal to the Rabi frequency $ \delta\xi $.
Values of the imaginary parts of frequencies $ E_r $ indicate that
only two pairs considerably participate in forming the long-time
photoelectron spectra $ I^{\rm lt} $. Moreover, the frequencies $
E_r $ of these two pairs nearly coincide [see Fig.~\ref{fig3}(a)].
This explains, why the photoelectron ionization spectra $ I^{\rm
lt} $ are composed only of two peaks. We note that a two-peak
structure is preserved even in the limit of weak optical pumping.
The Fano-like zero at frequency $ E=E_L $ can even be revealed for
$ \Omega \rightarrow 0 $, see Sec.~IV later. This distinguishes
spectral profiles of the autoionization system compared to the
Fano model [see Fig.~\ref{fig2}(a) in \cite{PerinaJr2011}] having
one central peak for weak optical pumping. Similarly as in the
Fano model, the peaks broaden with the increasing pumping
parameter $ \Omega $.

If the strength of direct ionization path is comparable with those
of two indirect ionization paths, the structure of long-time
photoelectron ionization spectra is much richer [see
Fig.~\ref{fig2}(b)]. We can identify two sharp nearly
symmetrically positioned peaks in graphs in Fig.~\ref{fig2}(b),
with the mutual distance roughly given by the Rabi frequency $
\delta\xi $. Contrary to two symmetrically positioned peaks
observed for large values of $ q_a $, $ q_b $ in
Fig.~\ref{fig2}(a), the widths of these peaks practically do not
depend on the pumping parameter $ \Omega $. Two additional much
broader peaks have been observed in the ionization spectra in
Fig.~\ref{fig2}(b) for larger values of pumping parameter $ \Omega
$. The first peak can be found in the area around $ E_L  $, the
second one occurs roughly one Rabi frequency $ \delta\xi $ on the
left-hand side of the first peak. If the pumping parameter $
\Omega $ increases, widths of both peaks and the distance between
the peaks constituting the Autler-Townes doublet increases.
Moreover, the analysis of behavior of eight complex frequencies $
E_r $ has shown that only two frequency pairs are located near the
real axis of complex frequency $ E $ and thus build the spectrum $
I^{\rm lt} $ [see Fig.~\ref{fig3}(b)]. This explains why the peaks
are mutually shifted roughly by one Rabi frequency. We note that
the behavior of the second two peaks resembles that found for the
two peaks in the long-time photoelectron spectrum of the
ionization system interacting with a neighbor in
\cite{PerinaJr2011} [compare Fig.~3(b) in \cite{PerinaJr2011}].

\begin{figure}  
 (a) \includegraphics[scale=0.4]{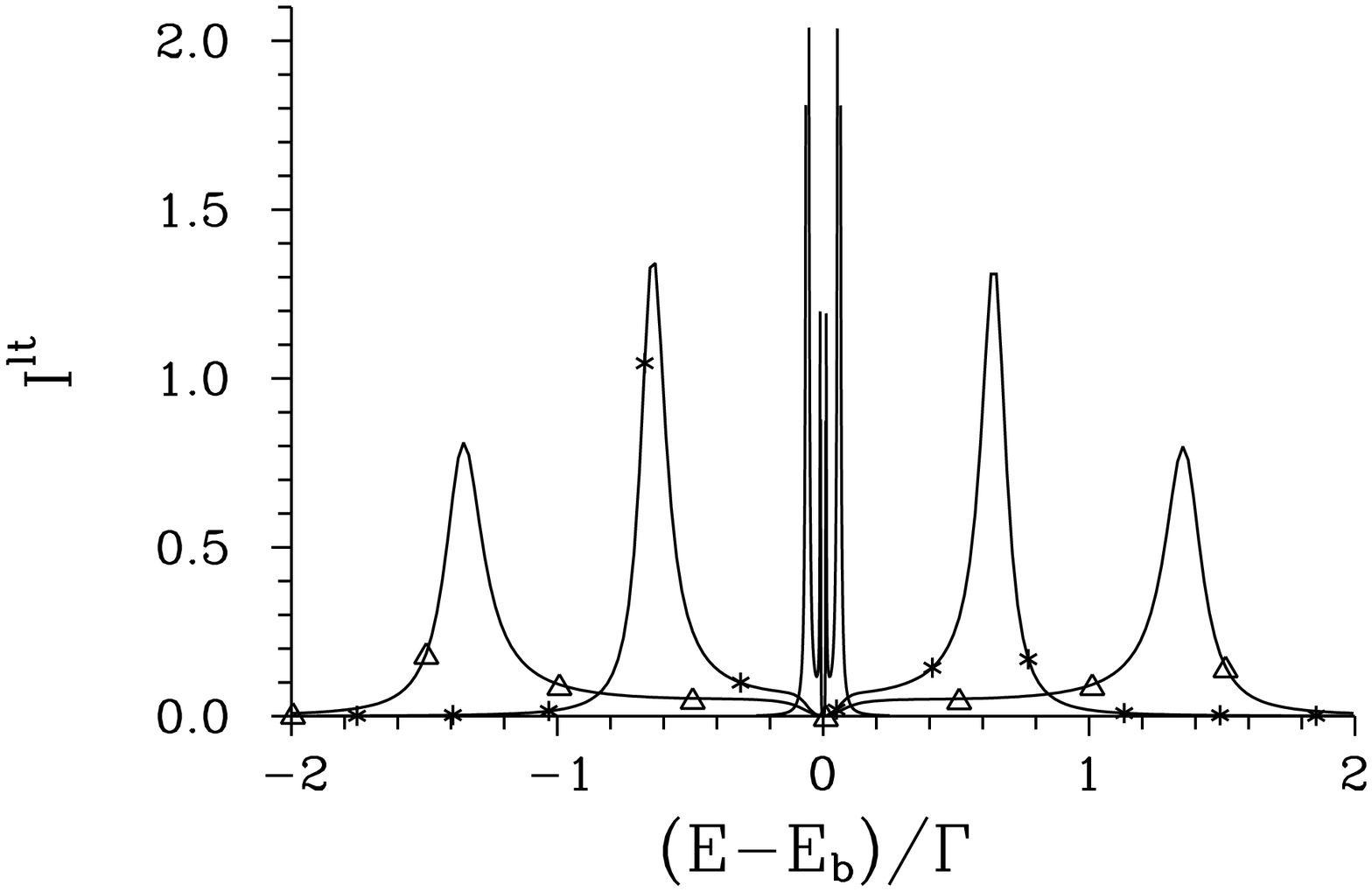}

 \vspace{7mm}
 (b) \includegraphics[scale=0.4]{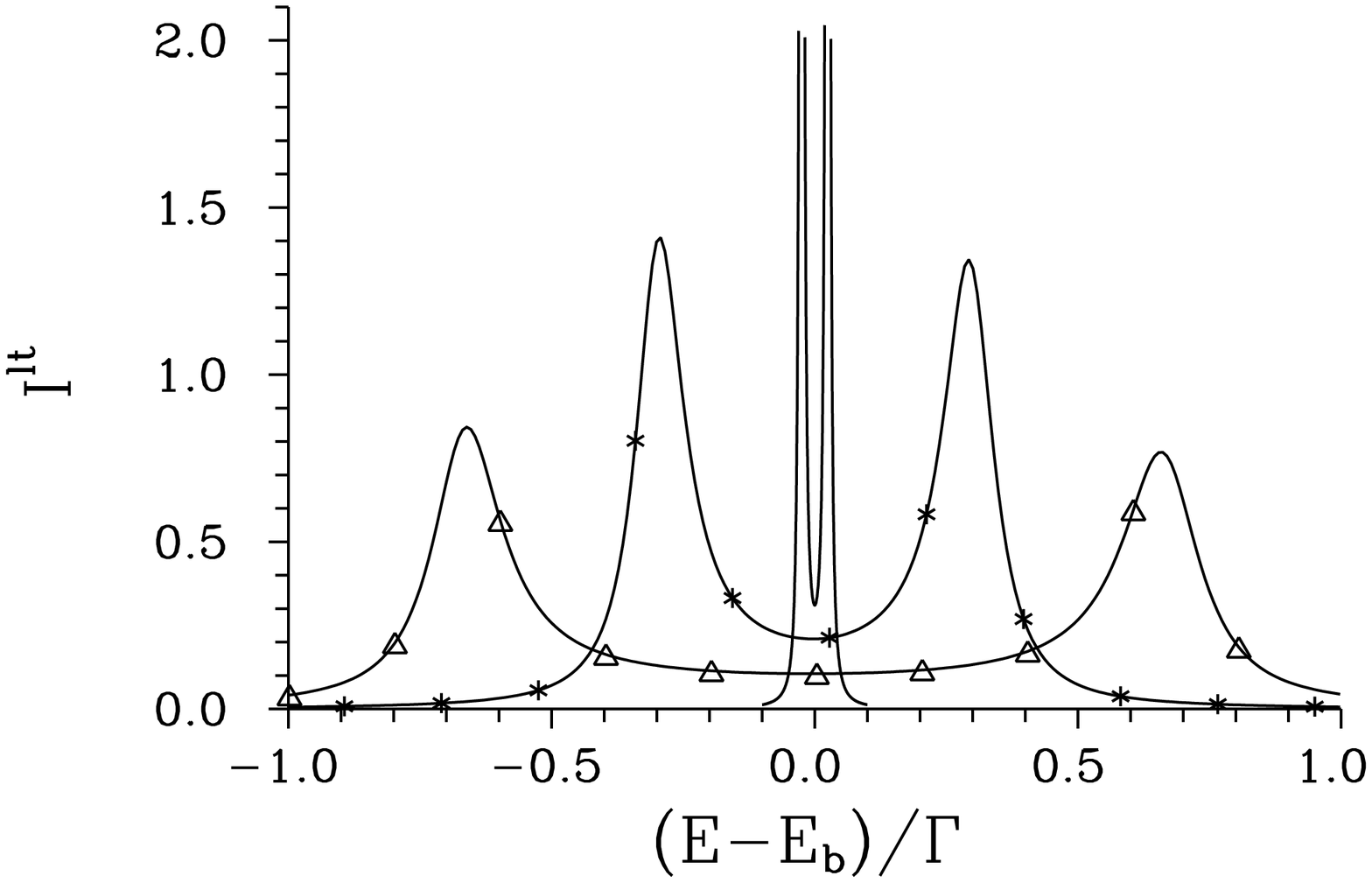}

 \caption{Long-time photoelectron ionization spectra $ I^{\rm lt} $ for
  (a) $ q_a = 100 $, $ q_b = 1 $ and (b) $ q_a= 1 $, $ q_b = 100 $ for different
  values of pumping parameter $ \Omega $: $ \Omega = 0.1
  $ (solid curve), $ \Omega = 1 $ (solid curve with $ \ast $),
  and $ \Omega = 2 $ (solid curve with $ \triangle $);
  $ \gamma_a = \gamma_b =1 $, $ E_a=E_b=E_L=1 $, $ J_{ab} = 0 $.}
\label{fig4}
\end{figure}

If indirect ionization including atom $ a $ prevails over the two
remaining ionization channels, the long-time photoelectron
ionization spectra are composed of two symmetrically positioned
peaks. Moreover, the probability to ionize the state with the
frequency $ E_a $ is practically zero [see Fig.~\ref{fig4}(a)].
The greater the pumping parameter $ \Omega $ is, the larger is the
distance between two peaks and the peaks are broader. This
behavior is qualitatively similar to that found for two side-peaks
in the spectra of the ionization system interacting with a
neighbor investigated in \cite{PerinaJr2011} [compare Fig.~2(b) in
\cite{PerinaJr2011}]. However, states with frequencies in the
middle of the spectrum are only weakly occupied due to the
interference with the additional ionization path coming through
the autoionizing state $ |1\rangle_b $.

When the autoionization exploiting state $ |1\rangle_b $ is
dominant, the Autler-Townes splitting of the photoelectron
ionization spectrum \cite{Autler1955} naturally occurs, as
documented in Fig.~\ref{fig4}(b). Even a weak indirect ionization
based on the presence of neighbor atom $ a $ is sufficient to
split the ionization peak typical of smaller values of the pumping
parameter $ \Omega $ into two symmetrically positioned side-peaks.
Symmetric two-peak photoelectron ionization spectra are thus
observed independently of the value of pumping parameter $ \Omega
$. The greater the pumping parameter $ \Omega $ is, the more
distant and broader are the peaks.

\begin{figure}  
 \includegraphics[scale=0.6]{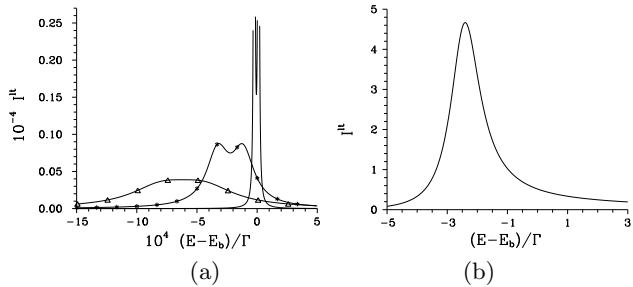}

 \hspace{5mm} (a) \hspace{30mm} (b)

 \caption{Long-time photoelectron ionization spectra $ I^{\rm lt} $ for
  (a) $ \Omega = 0.5 \times 10^{-2} $ (solid curve), $ \Omega = 3 \times 10^{-2} $
  (solid curve with $ \ast $), $ \Omega = 5 \times 10^{-2} $ (solid curve
  with $ \triangle $) and (b) $ \Omega = 1 $;
  $ q_a = 100 $, $ \gamma_a = 1 \times 10^{-4} $, $ q_b = \gamma_b =1 $,
  $ E_a=E_b=E_L=1 $, $ J_{ab} = 0 $.}
\label{fig5}
\end{figure}

In molecular condensates, the dipole-dipole interaction between
the neighbor molecules has typical energies 1 - 10~meV, whereas
energies in eV characterize the Coulomb configuration interaction.
The ratio $ \gamma_a / \gamma_b $ thus equals $ 10^{-4} - 10^{-6}
$ in this case. As the dipole-dipole interaction is weak ($
\gamma_a \ll 1 $), we are in the regime of $ q_a \gg 1 $ assuming
comparable values of the dipole moments $ \mu_a $ and $ \mu $. The
role of the state $ |1\rangle_a $ in forming the photoelectron
ionization spectra is important provided that the optical pumping
dipole and dipole-dipole interactions have comparable strengths.
This occurs only for weaker optical pumping. The needed pumping
amplitudes $ \alpha_L $ are thus by two or three orders of
magnitude lower compared to those used in typical ionization
experiments. An example of the dependence of the long-time
photoelectron ionization spectra on the pumping parameter $ \Omega
$ is shown in Fig.~\ref{fig5}. We can see in Fig.~\ref{fig5}(a)
that the presence of molecule $ a $ leads to splitting of the
spectral profile into two narrow peaks for smaller values of the
pumping parameter $ \Omega $. Widths of these peaks broaden, their
mutual distance increases and their central frequencies shift
towards the lower frequencies as the values of pumping parameter $
\Omega $ increase. However, the peaks are gradually absorbed into
an asymmetric spectral profile found for larger values of pumping
parameter $ \Omega $ [see Fig.~\ref{fig5}(b)] and being typical
for the Fano model with one autoionization level at molecule $ b
$. The role of molecule $ a $ in the formation of ionization
spectra is thus negligible for sufficiently large values of the
pumping parameter $ \Omega $. On the other hand, the interaction
with molecule $ a $ substantially modifies spectral profiles for
smaller values of pumping parameter $ \Omega $, as the comparison
of curves in Fig.~\ref{fig5} and those of Fig.~3(a) in
\cite{PerinaJr2011} clearly reveals.

\begin{figure}  
 \includegraphics[scale=0.4]{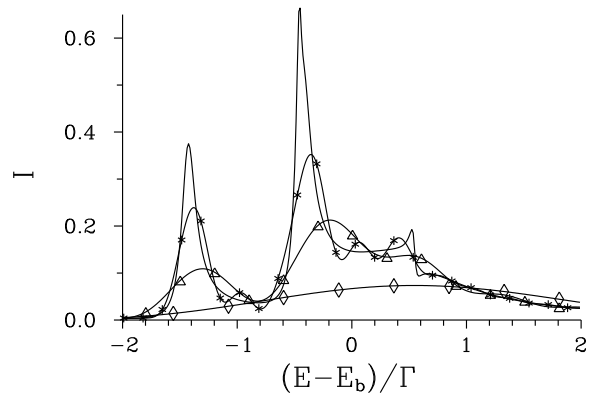}

 \caption{Photoelectron ionization spectra $ I $ at different times $ t $ [$ I(t) =
  I_0(t) + I_1(t) $]:
  $ t=1 $ (solid curve with $ \diamond $), $ t=5 $
  (solid curve with $ \triangle $), $ t=10 $
  (solid curve with $ \ast $), and $ t \rightarrow \infty $
  (solid curve). Spectrum $ I $ is determined as $ I(E,t) =
  |d_0(E,t)|^2 + |d_1(E,t)|^2 $; $ q_a = q_b =\gamma_a = \gamma_b =1 $,
  $ E_a=E_b=E_L=1 $, $ J_{ab} = 0 $, $ \Omega = 4 $.}
\label{fig6}
\end{figure}

Ionization spectra discussed up to now characterize the stationary
long-time limit. On using the formulas written in Sec.~II,
temporal aspects of forming the photoelectron ionization spectra
can also be addressed. As the time $ t $ increases, the initially
flat photoelectron ionization spectrum is gradually `focused' into
its long-time shape. Naturally, the creation of sharper features
in the long-time spectrum requires longer times, as illustrated in
Fig.~\ref{fig6}.

\section{Fano and Fano-like zeros in long-time photoelectron ionization spectra}

The Fano (Fano-like) and dynamical zeros belong to the most
important features of the long-time photoelectron ionization
spectra. The form of these spectra built from up to eight
Lorentzian curves [see Eqs.~(\ref{12}), (\ref{13}), and
(\ref{16})] allows to draw general conclusions about the number of
Fano and dynamical zeros (see Appendix B). It also allows to
develop a numerical method for finding frequencies of these zeros.
According to this analysis, no more than three Fano zeros can
exist.

Our investigations have revealed only one Fano zero (present for
arbitrarily strong pumping) under specific conditions. In the
limit of weak optical pumping, two Fano-like zeros have been
identified.

As for the Fano zero, its frequency $ E_F $ as well as conditions
for its observation can be revealed using a suitable canonical
transformation \cite{Luks2010}. From the physical point of view,
this Fano zero may occur only provided that completely destructive
interference between two ionization paths at the atom $ b $ occurs
as discovered in \cite{Fano1961}. In our model, we have two
additional ionization paths containing the energy transfer from
the excited state $ |1\rangle_a $ of the atom $ a $. In more
detail, the first path contains direct energy transfer ($ J $)
into the state $ |E_F) $, whereas the second additional path
includes energy transfer between the states $ |1\rangle_a $ and $
|1\rangle_b $ ($ J_{ab} $) and the Coulomb configurational
interaction ($ V $). These two additional ionization paths can
also cancel each other under suitable conditions. A detailed
analysis \cite{Luks2010} has shown that the state $ |E_F) $ in the
continuum is completely decoupled from both the states $
|0\rangle_b $ and $ |1\rangle_a $ only provided that
\cite{Luks2010}
\begin{equation}  
 \mu_b/\mu = J_{ab}/J .
\end{equation}
The frequency $ E_F $ of the observed Fano zero is then given as $
E_F = E_b - \gamma_b q_b $.

\begin{figure}  
 \includegraphics[scale=0.7]{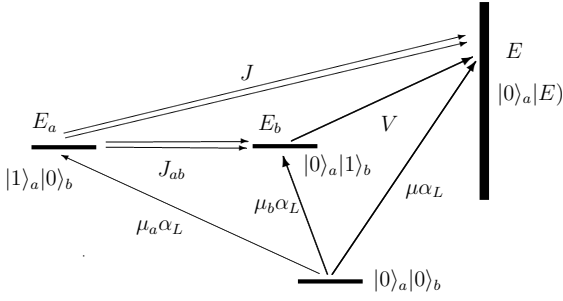}
 \caption{Scheme of the autoionization system $ b $ interacting with
  the two-level atom $ a $ valid for negligible
  probabilities of the states with two excited or ionized electrons.
  The state $ |0\rangle_a|0\rangle_b $ is the ground state, the state $ |0\rangle_a|1\rangle_b $
  ($ |1\rangle_a|0\rangle_b $) contains an excited electron at the atom $ b $ ($ a $), and
  the state $ |E)|0\rangle_a $ involves an ionized electron at the atom $ b $.}
\label{fig7}
\end{figure}

Two Fano-like zeros can be revealed in the limit of weak optical
pumping, i.e., when $ \mu_a\alpha_L $, $ \mu_b\alpha_L $, $ \mu
\alpha_L $ are much lower than $ V $, $ J $. In this limit, the
probability of having the states $ |1\rangle_a |1\rangle_b $ and $
|1\rangle_a |E) $ describing two excited or ionized electrons is
negligibly small compared to the other probabilities. A simplified
scheme of states as shown in Fig.~\ref{fig7} can then be
considered. This simplified model is suitable for the application
of a canonical transformation \cite{Fano1961} that gives two
Fano-like zeros. This transformation 'incorporates' the states $
|0\rangle_a|1\rangle_b $ and $ |1\rangle_a|0\rangle_b $ into the
states $ |E) $ of the continuum.

In this simplified scheme, we consider the Hamiltonian $
\hat{H}^{\rm oneexc}_{ab} $ that quantifies the energy of states
with one excited or ionized electron:
\begin{eqnarray}   
 \hat{H}^{\rm oneexc}_{ab} &=& E_a|1\rangle_a{}_a\langle1| + E_b|1\rangle_b{}_b\langle1|
  + \int dE \, E|E)(E| \nonumber \\
 & & \hspace{-7mm} \mbox{}  + \left[ J_{ab}|1\rangle_b
  {}_b\langle 0| |0\rangle_a {}_a\langle 1|
  + {\rm H.c.} \right] \nonumber \\
 & & \hspace{-7mm} \mbox{} +
  \int dE \, \left[ J |E)\, {}_b\langle 0| |0\rangle_a {}_a\langle 1|+\mbox{H.c.} \right]
  \nonumber \\
 & & \hspace{-7mm} \mbox{} +
  \int dE \, \left[ V |E)\,{}_b\langle 1| |0\rangle_a {}_a\langle 0| +\mbox{H.c.} \right].
\label{22}
\end{eqnarray}

We need to find eigenstates of the Hamiltonian $ \hat{H}^{\rm
oneexc}_{ab} $ and the corresponding dipole moments for the
transitions from the ground state $ |0\rangle_a|0\rangle_b $ into
the states arising from the diagonalization. We note that the
spectral long-time behavior of the system with the Hamiltonian $
\hat{H}^{\rm oneexc}_{ab} $ written in Eq.~(\ref{22}) has been
analyzed in \cite{Leonski1991} from the point of view of dc-field
coupling of two autoionizing levels.

The eigenstates $ |E)) $ of the Hamiltonian $ \hat{H}^{\rm
oneexc}_{ab} $ can be expressed as a linear superposition of the
states $ |1\rangle_a |0\rangle_b $, $ |0\rangle_a |1\rangle_b $,
and $ |0\rangle_a|E) $:
\begin{eqnarray}   
 |E)) &=& a(E) |1\rangle_a |0\rangle_b + b(E) |1\rangle_b
  |0\rangle_a  \nonumber \\
 & & \mbox{} + \int dE' \beta(E,E')|E') |0\rangle_a ,
\label{23}
\end{eqnarray}
where $ a(E) $, $ b(E) $, and $ \beta(E,E') $ are the coefficients
of the superposition. They obey the following system of linear
algebraic equations stemming from the stationary Schr\"{o}dinger
equation:
\begin{eqnarray}   
 & & E_a a(E) + J_{ab}^*b(E) \nonumber \\
 & & \hspace{2cm} \mbox{}  +  \int dE' J^*(E')\beta(E,E') = E a(E) ,
  \nonumber \\
 & & J_{ab} a(E) + E_b b(E) \nonumber \\
 & & \hspace{2cm} \mbox{} + \int dE' V^*(E')\beta(E,E') = E b(E) ,
  \nonumber \\
 & & J(E') a(E) + V(E') b(E) \nonumber \\
 & & \hspace{2cm} \mbox{} + E' \beta(E,E') = E \beta(E,E').
\label{24}
\end{eqnarray}

The third equation in (\ref{24}) can be solved in the following
form:
\begin{eqnarray}   
 \beta(E,E') &=& \frac{V(E')b(E)+J(E')a(E)}{E-E'+i\varepsilon }
  \nonumber \\
 & &  + F(E) \delta(E-E') ,
\label{25}
\end{eqnarray}
where $ \varepsilon > 0 $ ($ \varepsilon \rightarrow 0 $ is
assumed) and $ \delta $ means the Dirac $ \delta- $function. The
coefficient $ F(E) $ is determined from the normalization of state
$ |E)) $. The substitution of the solution in Eq.~(\ref{25}) into
the first two equations in (\ref{24}) gives the set of coupled
equations for the coefficients $ a(E) $ and $ b(E) $:
\begin{eqnarray}   
 \left[ \begin{array}{cc} \tilde{E}_a-i\gamma_a-E &
 \tilde{J}_{ab}^*-i\pi J^*(E)V(E) \\ \tilde{J}_{ab}-i\pi J(E)V^*(E)
 & \tilde{E}_b-i\gamma_b-E \end{array} \right] & & \nonumber \\
 & & \hspace{-7cm} \mbox{} \times \left[ \begin{array}{c} a(E) \\ b(E) \end{array} \right] = -
 \left[ \begin{array}{c} J^*(E) F(E) \\ V^*(E)
 F(E) \end{array} \right].
\label{26}
\end{eqnarray}
The damping constants $ \gamma_a $ and $ \gamma_b $ are given as $
\gamma_a = \pi|J(E_0)|^2 $ and $ \gamma_b = \pi|V(E_0)|^2 $ and
the frequency $ E_0 $ lies in the center of ionization spectrum.
The renormalized frequencies $ \tilde{E}_a $ and $ \tilde{E}_b $
and the coupling constant $ \tilde{J}_{ab} $ are determined along
the expressions:
\begin{eqnarray}   
 \tilde{E}_a(E) &=& E_a + {\cal P} \int dE'
  \frac{|J(E')|^2}{E-E'} , \nonumber \\
 \tilde{E}_b(E) &=& E_b + {\cal P} \int dE'
  \frac{|V(E')|^2}{E-E'} , \nonumber \\
 \tilde{J}_{ab}(E) &=& J_{ab} + {\cal P} \int dE'
 \frac{J(E')V^*(E')}{E-E'} ;
\label{27}
\end{eqnarray}
$ {\cal P} $ denotes the principal value. The solution of two
linear algebraic equations (\ref{26}) can be found, e.g., by
finding the inverse matrix of the system. The normalization
condition $ |a(E)|^2 + |b(E)|^2 + \int dE' |\beta(E,E')|^2 = 1 $
is fulfilled provided that we choose $ F(E) = 1 $. The
coefficients $ a(E) $, $ b(E) $, and $ \beta(E,E') $ can then be
derived in their final form:
\begin{eqnarray}    
 a(E) &=& \frac{ (E-\tilde{E}_b) J^*(E) +
  \tilde{J}_{ab}^*V^*(E) }{ {\cal D}(E) } , \nonumber \\
 b(E) &=& \frac{ (E-\tilde{E}_a) V^*(E) +
  \tilde{J}_{ab} J^*(E) }{ {\cal D}(E) } , \nonumber \\
 \beta(E,E') &=& \frac{V(E')b(E)+J(E')a(E)}{E-E'+i\varepsilon }
  \nonumber \\
 & & \mbox{}  + \delta(E-E') .
\label{28}
\end{eqnarray}
The symbol $ {\cal D}(E) $ in Eq.~(\ref{28}) stands for the
determinant of the matrix in Eq.~(\ref{26}) that is given as:
\begin{eqnarray}  
 {\cal D}(E) &=& (E-\tilde{E}_a) (E-\tilde{E}_b) + i\gamma_a
  (E-\tilde{E}_b) \nonumber \\
 & & \mbox{} \hspace{-1cm} + i\gamma_b (E-\tilde{E}_a) -
  |\tilde{J}_{ab}|^2 +2i{\rm Re}\{t(E)\}.
\label{29}
\end{eqnarray}
In Eq.~(\ref{29}), $ t(E) = \pi \tilde{J}_{ab}^* J(E) V^*(E) $.
The function $ t(E) $ is nonzero only if the energy transfer
between the excited states $ |1\rangle_a $ and $ |1\rangle_b $
occurs.

The optical-field interaction between the ground state $
|0\rangle_a |0\rangle_b $ and the states with one excited or
ionized electron as described by the Hamiltonian $ \hat{H}^{\rm
oneexc}_{ab} $ is governed by the Hamiltonian $ \hat{H}_{ab}^{\rm
opt} $:
\begin{eqnarray}   
 \hat{H}^{\rm opt}_{ab} &=&
  \left[ \mu_a \alpha_L \exp(-iE_Lt)|1\rangle_a {}_a\langle0|
   |0\rangle_b {}_b\langle0| + \mbox{H.c.}
  \right] \nonumber \\
  & & \hspace{-4mm} \mbox{} + \left[ \mu_b \alpha_L \exp(-iE_Lt) |1\rangle_b {}_b\langle0|
   |0\rangle_a {}_a\langle0| + \mbox{H.c.}
  \right] \nonumber \\
 & & \hspace{-7mm} \mbox{} +
  \int dE \, \left[ \mu \alpha_L \exp(-iE_Lt) |E)\, {}_b\langle0|
   |0\rangle_a {}_a\langle0| +\mbox{H.c.}
  \right] . \nonumber \\
 & &
\label{30}
\end{eqnarray}
The Hamiltonian $ \hat{H}_{ab}^{\rm opt} $ transformed into the
basis $ |0\rangle_a |0\rangle_b $ and $ |E)) $ reads:
\begin{eqnarray}   
 \hat{H}^{\rm opt}_{ab} &=&
  \int dE \left[ \bar{\mu}(E) \alpha_L \exp(-iE_Lt) |E)) \,
  {}_a\langle0| {}_b\langle0|  +\mbox{H.c.}
  \right] . \nonumber \\
 & &
\label{31}
\end{eqnarray}
The dipole moment $ \bar{\mu} $ in the transformed basis is given
by the formula
\begin{equation}   
 \bar{\mu}(E) = \mu_a a^*(E) + \mu_b b^*(E) + \int dE' \mu(E')
  \beta^*(E,E').
\label{32}
\end{equation}
This formula can be recast into the following form:
\begin{eqnarray}   
 \bar{\mu}(E) &=& \left[ 1 + \frac{ (q_a+i) [\gamma_a(E-\tilde{E}_b) + t^*]
  }{{\cal D}^*(E) } \right. \nonumber \\
 & & \hspace{-3mm} \left. \mbox{} +  \frac{ (q_b+i)[\gamma_b(E-\tilde{E}_a) + t] }{{\cal D}^*(E) }
  \right] \mu(E).
\label{33}
\end{eqnarray}
We remind that the parameters $ q_a $ and $ q_b $ are defined in
the caption to Fig.~\ref{fig2}.

The frequencies $ E_F $ giving the positions of Fano-like zeros
can be easily identified using the condition $ \bar{\mu}(E_F) = 0
$ in Eq.~(\ref{33}). This leads to the quadratic equation:
\begin{eqnarray}   
 & & \hspace{-10mm}  [E-\tilde{E}_b]^2 + [-\tilde{E}_a + \tilde{E}_b
  + q_a\gamma_a + q_b\gamma_b] [E-\tilde{E}_b] \nonumber \\
 & & \hspace{-3mm} \mbox{}  + [q_a t^*+q_b(t-\gamma_b\tilde{E}_a + \gamma_b\tilde{E}_b ) -
  |\tilde{J}_{ab}|^2 ] = 0 .
\label{34}
\end{eqnarray}
The solution of quadratic equation (\ref{34}) reveals the
frequencies $ E_F $ of at most two Fano zeros:
\begin{equation}   
 \left[ E_F \right]_{1,2} = \frac{\tilde{E}_a + \tilde{E}_b - q_a\gamma_a -
   q_b\gamma_b}{2} \pm \frac{\sqrt{D}}{2}.
\label{35}
\end{equation}
They occur for a non-negative discriminant $ D $,
\begin{eqnarray}   
 D &=& (\tilde{E}_a - \tilde{E}_b)^2 + (q_a\gamma_a +
  q_b\gamma_b)^{2} - 2(\tilde{E}_a - \tilde{E}_b)
  \nonumber \\
 & & \mbox{} \times (q_a\gamma_a - q_b\gamma_b) +
  4|\tilde{J}_{ab}|^2 - 4q_a t^* - 4q_b t.
\label{36}
\end{eqnarray}
We note that the frequencies $ E_F $ of two Fano zeros written in
Eq.~(\ref{36}) coincide with those revealed in \cite{Leonski1987}
for a double Fano system provided that $ \tilde{J}_{ab} = 0 $.

\section{Dynamical zeros in long-time photoelectron ionization spectra}

\begin{figure}  
 (a) \includegraphics[scale=0.4]{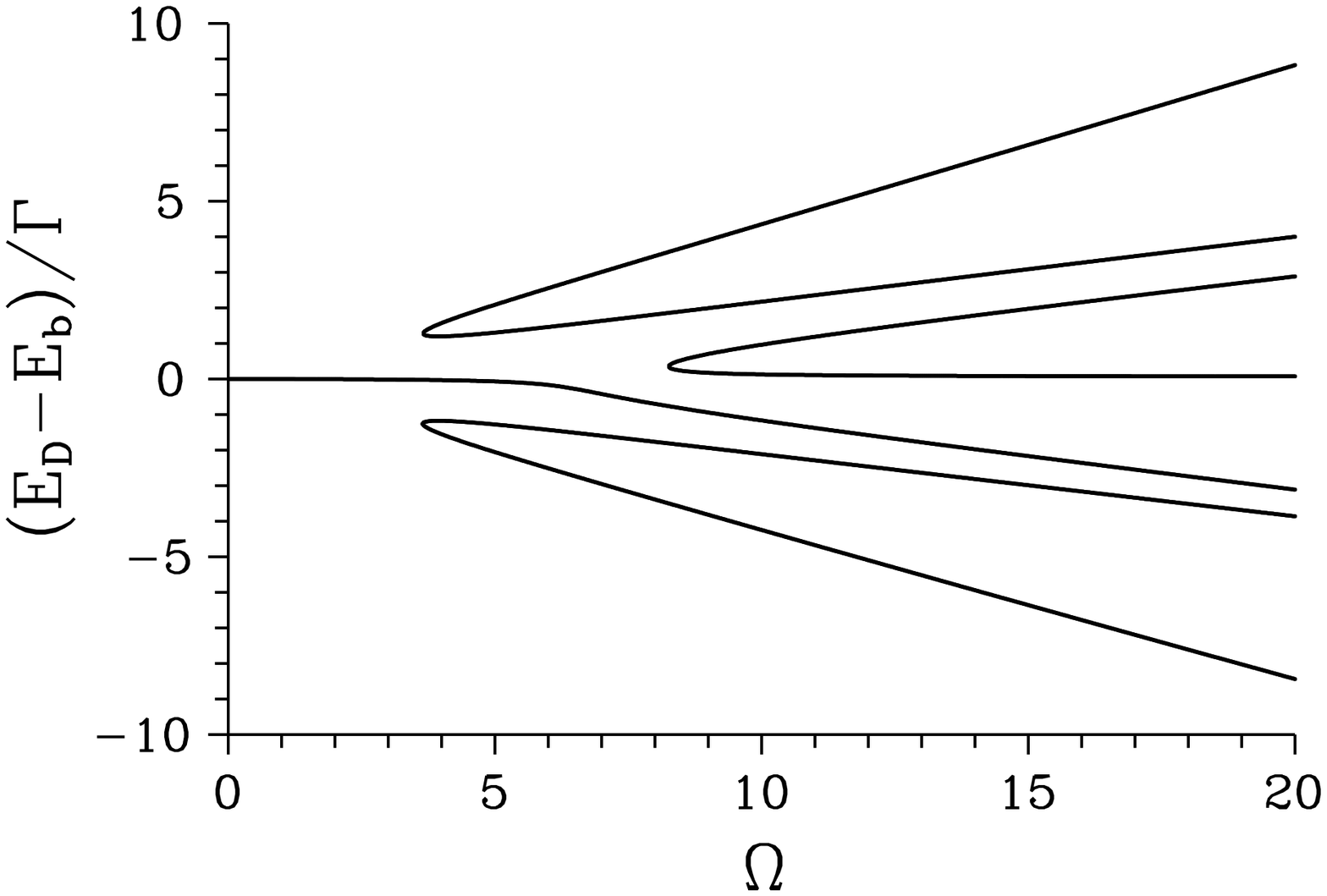}

 \vspace{7mm}
 (b) \includegraphics[scale=0.4]{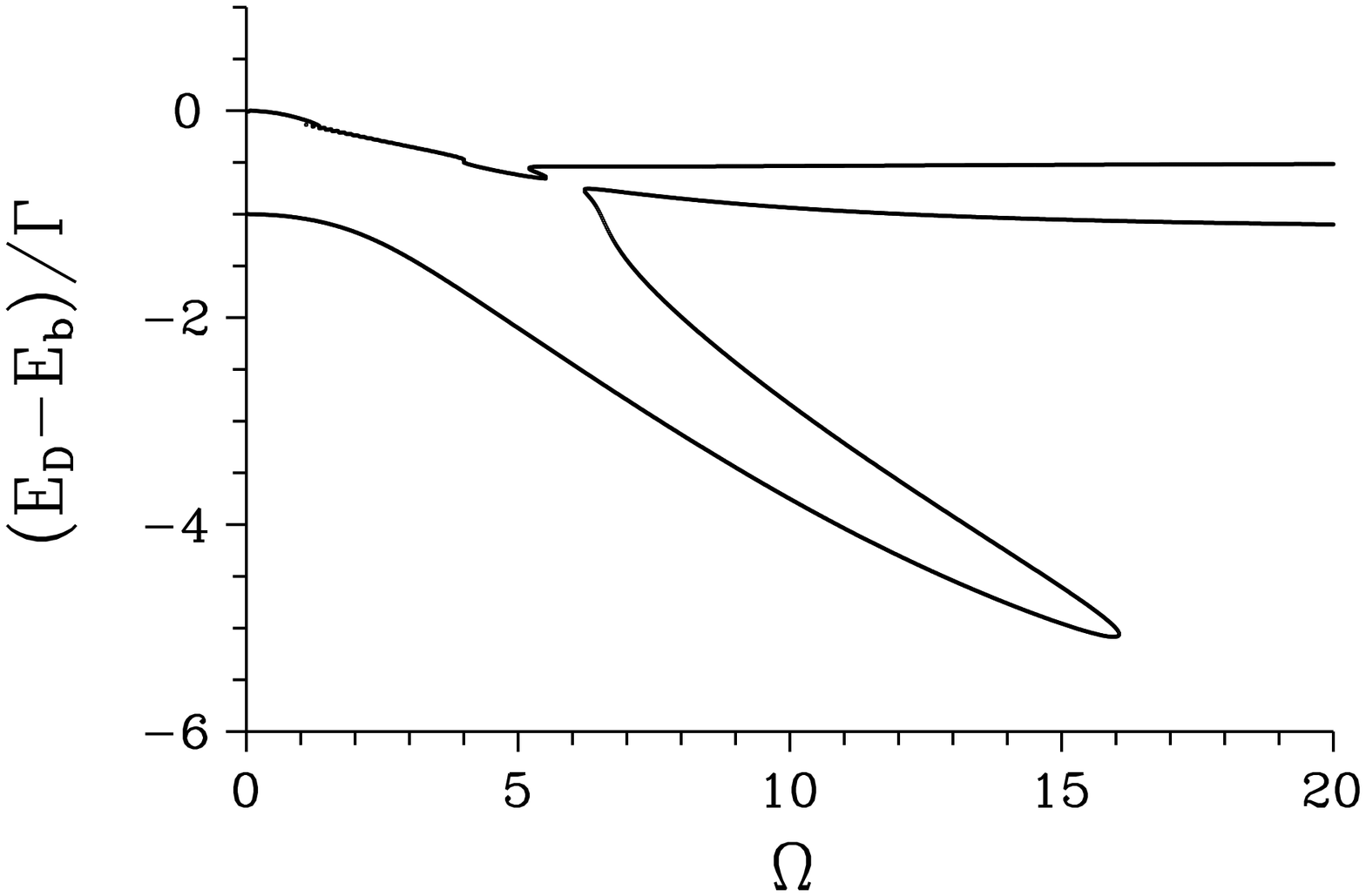}

 \caption{Normalized frequencies $ (E_D-E_b)/\Gamma $ of dynamical zeros
  as they depend on pumping parameter $ \Omega $ for resonant pumping of atom $ a $:
  (a) $ q_a = q_b = 100 $, (b) $ q_a = q_b = 1 $; $ \gamma_a = \gamma_b = 1 $,
  $ E_a = E_b = E_L = 1 $, $ J_{ab} = 0 $.}
\label{fig8}
\end{figure}

\begin{figure}  
 \includegraphics[scale=0.4]{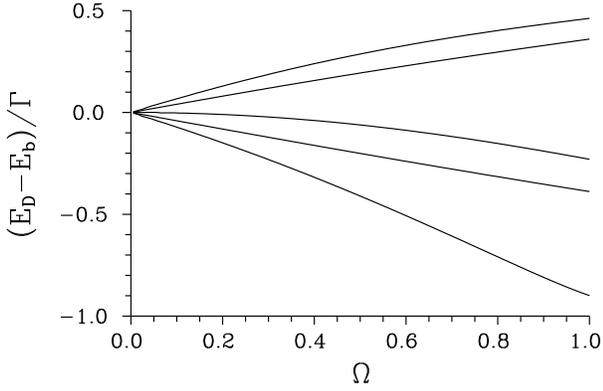}

 \caption{Normalized frequencies $ (E_D-E_b)/\Gamma $ of dynamical zeros
  as they depend on pumping parameter $ \Omega $ for resonant pumping of atom $ a $:
  values of parameters are written in the caption to Fig.~\ref{fig5}.}
\label{fig9}
\end{figure}

The analysis contained in Appendix~B reveals that up to 15
dynamical zeros can exist. Their frequencies $ E_D $ can be
determined along the recipe described in Appendix~B. We should
note that frequencies $ E_D $ of two of the dynamical zeros
coincide with the frequencies $ E_F $ of the Fano-like zeros
discussed in Sec.~IV in the limit of weak optical pumping. If we
plot the normalized frequencies $ (E_D-E_b)/\Gamma $ of dynamical
zeros as functions of the pumping parameter $ \Omega $, we arrive
at graphs similar in shape to those appearing in the model of
ionization system interacting with a neighbor (see Fig.~6 in
\cite{PerinaJr2011}). Graphs obtained for the analyzed model are
in general more complex. However, creation and annihilation of
dynamical zeros in pairs represents their most typical feature. As
an example, the graphs corresponding to the values of parameters
defined in the caption to Fig.~\ref{fig2} are shown in
Fig.~\ref{fig8}. They demonstrate the 'creation' and
'annihilation' of dynamical zeros is pairs. However, dynamical
zeros can emerge even in greater numbers. This is documented in
Fig.~\ref{fig9} where even five dynamical zeros occur with
frequency $ E_D = E_b $ for $ \Omega = 0 $ for values of
parameters typical for molecular condensates. Similarly as in the
long-time photoelectron ionization spectra of the ionization
system interacting with a neighbor \cite{PerinaJr2011}, splitting
of the normalized frequencies $ (E_D-E_b)/\Gamma $ appropriate to
the spectra $ I^{\rm lt}_0 $ and $ I^{\rm lt}_1 $ is observed for
non-resonant pumping of the two-level atom $ a $ ($ E_a \neq E_L
$). The symmetry $ \Omega \leftrightarrow -\Omega $ mentioned in
\cite{PerinaJr2011} is also preserved in the analyzed
autoionization system.

\section{Conclusions}

The long-time photoelectron ionization spectra of an
autoionization system interacting with a neighbor two-level atom
have been investigated in several distinct regimes. They typically
consist of several peaks with central positions and spectral
widths depending on the pumping strength. As a consequence of
interference of several ionization paths, zeros in the long-time
photoelectron ionization spectral profiles occur. Whereas only one
genuine Fano zero has been found under special conditions, two
Fano-like zeros observed only for weak optical pumping have been
identified for a general system constructing a suitable canonical
transformation. The long-time photoelectron ionization spectra
conditioned by the state of the neighbor two-level atom exhibit
permanent Rabi oscillations. Spectral dynamical zeros observed
once in the Rabi period have been revealed in these spectra. The
frequencies of these dynamical zeros depending on the strength of
optical pumping as well as the projected state of the neighbor
two-level atom have been analyzed.

\acknowledgments Support by the projects 1M06002, COST OC 09026,
and Operational Program Research and Development for Innovations -
European Social Fund (project CZ.1.05/2.1.00/03.0058) of the
Ministry of Education of the Czech Republic as well as the project
IAA100100713 of GA AV \v{C}R is acknowledged.

\appendix

\section{Determination of poles of the Lorentzian curves giving photoelectron spectra}

The complex frequencies $ E_r $ giving poles of the Lorentzian
curves composing photoelectron ionization spectra are determined
as a sum of eigenvalues $ \Lambda_{M^e,j} $ of the matrix $ {\bf
M^e} $ ($ j=1,\ldots,4 $) and frequencies $ \xi_{k} $ ($ k=1,2 $)
giving the Rabi oscillations of two-level atom $ a $; $ E_r =
\Lambda_{M^e,j} + \xi_{k} $. Whereas the frequencies $ \xi_{k} $
are written in Eq.~(\ref{11}), the eigenvalues $ \Lambda_{M^e,j} $
can be derived from roots of the fourth-order polynomial written
in the shifted frequency $ \tilde{\Lambda}_{M^e}$; $ \Lambda_{M^e}
= \tilde{\Lambda}_{M^e} - i\pi|\mu\alpha_L|^2 $:
\begin{equation}   
  [\tilde{\Lambda}_{M^e}]^4 + \alpha_3 [\tilde{\Lambda}_{M^e}]^3 + \alpha_2
   [\tilde{\Lambda}_{M^e}]^2 + \alpha_1 \tilde{\Lambda}_{M^e} + \alpha_0 = 0 .
\label{A1}
\end{equation}
The coefficients $ \alpha_j $ introduced in Eq.~(\ref{A1}) can be
derived from the elements of matrix $ {\bf M^e} $:
\begin{eqnarray}   
 \alpha_0 &=& (\Delta E_a + {\cal E}_b) \left[ -{\cal E}_b M_a M_a^c +
  (-\Delta E_a +i\gamma_a) M_b M_b^c \right. \nonumber \\
 & & \left. \mbox{}+ (M_a M_b^c j_{ab} + M_a^c M_b
  j_{ab}^c) \right] |\alpha_L|^2 + \left[ M_a M_a^c |\mu_a|^2  \right. \nonumber \\
 & & \left. \mbox{} + M_b^2 M_b^{c2} - (M_a \mu_a^* + M_a^c\mu_a) M_b M_b^c \right]
  |\alpha_L|^4 , \nonumber \\
 \alpha_1 &=& (-\Delta E_a +i\gamma_a)(\Delta E_a + {\cal E}_b){\cal
  E}_b + \left[ (\Delta E_a - i\gamma_a) |\mu_a|^2 \right. \nonumber \\
 & & \mbox{}  + (\Delta E_a +2{\cal E}_b) M_a M_a^c
  + (2\Delta E_a - i\gamma_a +2{\cal E}_b) \nonumber \\
 & & \left. \mbox{} \times M_b M_b^c \right]
  |\alpha_L|^2 + (\Delta E_a + {\cal E}_b) j_{ab} j_{ab}^c  \nonumber \\
 & & \mbox{} - \left[ (\mu_a+M_a) M_b^c j_{ab} + (\mu_a^* + M_a^c) M_b j_{ab}^c
  \right] |\alpha_L|^2,  \nonumber \\
 \alpha_2 &=& \Delta E_a (\Delta E_a -i\gamma_a) + (3\Delta E_a
  -2i\gamma_a) {\cal E}_b + {\cal E}_b^2 \nonumber \\
 & & \mbox{} - \left[ M_a M_a^c +
  |\mu_a|^2 + 2 M_b M_b^c \right] |\alpha_L|^2 - j_{ab} j_{ab}^c,
  \nonumber \\
 \alpha_3 &=& -2\Delta E_a - 2{\cal E}_b + i\gamma_a  .
\label{A2}
\end{eqnarray}
In Eq.~(\ref{A2}), $ {\cal E}_b = \Delta E_b - i\gamma_b + i\pi
|\mu\alpha_L|^2 $, $ M_a = \mu_a - i\pi\mu J^* $, $ M_a^c =
\mu_a^* - i\pi\mu^* J $, $ M_b = \mu_b - i\pi\mu V^* $, $ M_b^c =
\mu_b^* - i\pi\mu^* V $, $ j_{ab} = J_{ab} -i\pi J V^* $, and $
j_{ab}^c = J_{ab}^* -i\pi J^* V $. Roots of the polynomial in
Eq.~(\ref{A1}) can be found analytically, in principle, which
might be useful in special cases.

\section{Determination of Fano and dynamical zeros}

The specific form of the long-time solution for the photoelectron
spectra written in Eqs.~(\ref{12}), (\ref{13}), and (\ref{16})
allows to reformulate the condition in Eq.~(\ref{19}) for the
frequencies $ E_F $ of Fano zeros. They can be found as a common
solution of the following four equations:
\begin{equation}   
 \sum_{j=1}^{4} \frac{A_j^{kl} }{ E - \Lambda_{M^e,j} -\xi_l } = 0;
 \hspace{5mm} k=0,1, \hspace{2mm}  l=1,2.
\label{B1}
\end{equation}
The coefficients $ A_j^{kl} $ can be derived from the solution in
Eq.~(\ref{13}) as follows:
\begin{equation}  
 A_j^{kl} = \left[ {\bf K_l B^{\bf e\dagger} P^e}\right]_{kj}
  \left[ {\bf P}^{\bf e -1} {\bf c}(0) \right]_{j} .
\label{B2}
\end{equation}
Equations in (\ref{B2}) can be recast into the form of the
third-order polynomials $ p_{kl}(E) $:
\begin{equation}   
 p_{kl}(E) = \sum_{j=0}^{3} \alpha_j^{kl} E^j = 0
\label{B3}
\end{equation}
with the coefficients defined as:
\begin{eqnarray}   
 \alpha_0^{kl} &=& - \sum_{j=1}^{4} A_j^{kl} \prod_{m=1,m\neq j}^{4}
  (\Lambda_{M^e,m} + \xi_l) , \nonumber \\
 \alpha_1^{kl} &=& \sum_{j=1}^{4} A_j^{kl} \sum_{m=1, m\neq j }^{4}
  (\Lambda_{M^e,m} + \xi_l) \nonumber \\
  & & \mbox{} \times  \sum_{n=1,n\neq j,m}^{4}
  (\Lambda_{M^e,n} + \xi_l) , \nonumber \\
 \alpha_2^{kl} &=& - \sum_{j=1}^{4} A_j^{kl} \sum_{m=1,m\neq j}^{4}
  (\Lambda_{M^e,m} + \xi_l), \nonumber \\
 \alpha_3^{kl} &=& \sum_{j=1}^{4} A_j^{kl} .
\label{B4}
\end{eqnarray}
A Fano zero is identified only provided that its frequency $ E_F $
is found simultaneously among three roots of the polynomials $
p_{kl} $ for all $ k=0,1 $ and $ l=1,2 $. Moreover, the
corresponding root has to be real. In general, no more than three
Fano zeros can be found.

The third-order polynomials $ p_{kl} $ introduced in
Eq.~(\ref{B3}) are also useful in expressing the conditions $
|d_k^{\xi_1}(E,t)| = |d_k^{\xi_2}(E,t)| $ [equivalent to those
written in Eq.~(\ref{20})] for the occurrence of dynamical zeros
in the spectra $ I^{\rm lt}_k $, $ k=0,1 $:
\begin{eqnarray}   
 & & \hspace{-5mm} |p_{k1}(E)| \prod_{j=1}^{4} |E-\Lambda_{M^e,j} -\xi_2| \nonumber \\
 & &  = |p_{k2}(E)| \prod_{j=1}^{4} |E-\Lambda_{M^e,j}-\xi_1| ,
  \hspace{3mm} k=0,1.
\label{B5}
\end{eqnarray}
Equation (\ref{B5}) represents a fifteenth-order polynomial which
can have complex coefficients. If a root of this polynomial is
real, it gives the frequency $ E_D $ of a dynamical zero. In
principle, up to 15 dynamical zeros might exist.

\bibliography{perina}
\bibliographystyle{apsrev}

\end{document}